# Electrically Tunable Terahertz Chirality from Quantum Geometry

**Sobhan Subhra Mishra[1,2], Thomas CaiWei Tan [1,2], Faxian Xiu[3,4,5*], Ranjan Singh[6*]**

[1]*Division of Physics and Applied Physics, School of Physical and Mathematical Sciences, Nanyang Technological University, Singapore 637371*
[2]*Centre for Disruptive Photonic Technologies, Nanyang Technological University, Singapore 639798*
[3]*State Key Laboratory of Surface Physics and Department of Physics, Fudan University, Shanghai 200433, China*
[4]*Institute for Nanoelectronic Devices and Quantum Computing, Fudan University, Shanghai 200433, China*
[5]*Shanghai Research Center for Quantum Sciences, Shanghai, 201315, China*
[6]*Department of Electrical Engineering, University of Notre Dame, Notre Dame, IN, USA*

*\* Corresponding Authors- faxian@fudan.edu.cn, rsingh3@nd.edu*

## Abstract

Quantum geometry encoded in the momentum space structure of electronic wavefunctions, governs charge dynamics through Berry curvature, enabling unconventional transport and optical responses. In topological semimetals, this geometry is sampled over Fermi pockets, suggesting electrical control by Fermi surface tuning, yet such control has remained largely limited to DC transport. Here we show that electrostatic gating of the 3D Dirac semimetal $Cd_3As_2$ reshapes Fermi pockets surrounding photoinduced Floquet Weyl nodes, enabling electrical control of terahertz (THz) emission chirality. Gate tuning selectively modulates the Berry curvature driven linearly polarized THz component by up to 60% and 49% at positive and negative bias, respectively, while the orthogonal linearly polarized photon-drag component remains unchanged. With the two orthogonal fields intrinsically phase-locked at $\pi/2$ by the excitation geometry, the selective gate-tuned amplitude control enables the polarization tuning across the Poincaré sphere, achieving near-circular polarization ($\chi \approx -42°$) at +10 V. These results establish Fermi surface tuning as a general route to programmable quantum geometric control of chiral terahertz emission.

## Introduction

Quantum geometry, which characterizes the momentum space structure of electronic Bloch wavefunctions, provides a unified framework for understanding emergent phenomena in quantum materials[1,2]. It is naturally encoded in a complex geometric tensor whose real and imaginary parts capture distinct physical properties. The imaginary part corresponds to the Berry curvature and governs the anomalous velocity of charge carriers, giving rise to transverse transport responses. The real part defines the quantum metric and quantifies the distance between neighbouring quantum states, thereby controlling the spatial extent of Wannier orbitals and the strength of interband transitions[3,4]. Together these quantities form the quantum geometric tensor[1] $\mathcal{Q}_{\mu\nu} = g_{\mu\nu} + \frac{i}{2}\Omega_{\mu\nu}$, where $g_{\mu\nu}$ is the Fubini-Study metric and $\Omega_{\mu\nu}$ is the Berry curvature tensor. A broad class of nonlinear optical and transport responses ranging from the anomalous Hall effect[5–7] to the circular photogalvanic effect[8,9] descend from the Berry curvature component of this tensor. In Weyl semimetals, each nodal point acts as a quantised monopole source of Berry curvature in momentum space[10], and the net flux enclosed by the Fermi pocket directly determines the magnitude and sign of the anomalous velocity that drives transverse photocurrents. The Fermi pocket, as the constant energy surface enclosing these monopoles, is therefore a fundamental control parameter for quantum geometry mediated responses[11]. Reshaping the pocket changes the Berry curvature integral sampled by the charge carriers and, in turn, every transport coefficient that descends from it.

Electrical gating offers a uniquely reversible and continuous means of reshaping Fermi pockets in thin film geometries, without perturbing the crystal symmetry or the optical excitation conditions. Demonstrating that gate driven Fermi pocket reshaping quantitatively controls a Berry curvature mediated observable would establish a general strategy for programming the quantum geometry governed responses, including the photogalvanic effect, anomalous Hall effect, and nonlinear Hall effect[12,13], all of which share the same Berry curvature integral over the Fermi surface as their mathematical origin. However, while DC transport experiments have probed this connection in equilibrium, the principle has not been extended to dynamic optical observables.

Cd$_3$As$_2$ provides an ideal platform for such investigations due to its unique electronic properties[14–17]. Its pair of symmetry-protected Dirac nodes can be split into Weyl nodes of opposite chirality when time reversal symmetry is broken by circularly polarized photoexcitation, dressing the bands into Floquet Weyl states[18–20]. Each resulting Weyl node carries a Berry monopole charge $C = \pm 1$ and produces a curvature flux that diverges as $\Omega \sim C/2k^2$ near the nodal point[10,21], placing intense quantum geometric flux precisely at the Fermi pocket boundary. A small pocket tightly enclosing the node samples this intense curvature at the monopole core and yields a strong anomalous photocurrent. A larger pocket progressively incorporates regions of weaker curvature, diluting the Berry curvature integral and diminishing the response. Importantly, the Fermi level in Cd$_3$As$_2$ thin films can be continuously tuned by electrostatic gating[22,23], making it possible to electrically resize the Fermi pockets around the Weyl nodes and thereby control the quantum geometric contribution to the photocurrent along each crystallographic direction independently. Additionally, under oblique incidence, a circular photon drag contribution emerges[24,25] whose amplitude is set solely by photon momentum and is independent of Berry curvature. This mechanistic separation enables gate induced Fermi pocket reshaping to selectively tune one component, converting Berry flux modulation into programmable terahertz polarization control.

Here, we demonstrate that Fermi pocket manipulation in Cd$_3$As$_2$ enables all electrical control of the THz polarization state. By applying positive and negative gate voltages to a gated Cd$_3$As$_2$ thin film under fixed circularly polarized photoexcitation at oblique incidence, we tune the Fermi pocket size relative to the Floquet Weyl nodes. The Y polarized THz component is modulated by over 60% under positive bias and 49% under negative bias, while the X polarized component arising from the circular photon drag effect remains unchanged. Sweeping the gate voltage tunes the ratio of these two orthogonal THz fields. Since their intrinsic phase relationship is fixed by the excitation geometry, the polarization state of the emitted THz radiation is governed entirely by this electrically tunable amplitude ratio. At a specific gate voltage of 10 V, the two orthogonal amplitudes become equal, and the intrinsic 90° phase difference transforms the emission into a circularly polarized THz pulse. This electrically driven transition from elliptical to circular polarization is reversible and operates at room temperature. Our results establish the Fermi pocket size as a quantitative tuning parameter for quantum geometry mediated responses in topological semimetals and open a route toward gate programmable chiral THz sources for spectroscopy, imaging, and communications.

## Result and Discussion

Figure 1(a) illustrates the Floquet band engineering underlying our approach. In equilibrium, $Cd_3As_2$ hosts a pair of symmetry-protected Dirac nodes along the crystallographic c axis[14–16,26]. Circularly polarized femtosecond photoexcitation breaks time-reversal symmetry and splits each node into two Weyl nodes of opposite chirality separated by $\Delta k$ in momentum space.[18–20,27,28] Each Weyl node acts as a Berry curvature monopole with topological charge $C = \pm 1$, generating a curvature field that diverges as $\Omega \sim C/2k^2$ near the nodal point[10]. The photocurrent response of this Floquet-Weyl state contains two physically distinct contributions when pumped with circularly polarized light at oblique incidence (pump fluence = $1.02 \ mJ/cm^2$). All other contributions towards THz emission can be safely ignored when pumped with circularly polarized light as discussed in our previous work[20] and supplementary section S3. The first is anomalous photocurrent arising from asymmetric interband excitations weighted by the Berry curvature of the occupied Bloch states. The anomalous photocurrent density along the transverse (Y) direction is governed by the Berry curvature flux enclosed within the Fermi pocket[5,9,29] as

$$J_Y^A \propto \int_{pocket} \Omega(\mathbf{k}) \left[ f(\mathbf{k}) - f'(\mathbf{k}) \right] d^3k \qquad (1)$$

where $f(\mathbf{k})$ and $f'(\mathbf{k})$ are the Fermi-Dirac distributions of the initial and final states, and the integration extends over the constant-energy surface defined by $E_F$. Equation (1) is the Berry curvature component of the broader quantum geometric tensor $Q_{\mu\nu}^{(n)} = g_{\mu\nu}^{(n)} + \frac{i}{2}\Omega_{\mu\nu}^{(n)}$, where $g_{\mu\nu}^{(n)}$ is the quantum metric of band n. Near the Weyl node in $Cd_3As_2$, the Berry curvature contribution dominates the photocurrent response because the monopole divergence $\Omega \sim 1/k^2$ greatly exceeds the quantum metric contribution at the relevant Fermi energies. In this regime, the anomalous photocurrent couples exclusively to the antisymmetric Berry curvature component of the quantum geometric tensor, as the symmetric quantum metric component does not contribute to transverse Hall-type photocurrents, making the measured THz emission a direct, isolated probe of the Berry curvature geometry. Since the Berry curvature is concentrated near the Weyl node and falls off as $1/k^2$, the value of this integral depends on the Fermi pocket size. A small pocket tightly enclosing the node samples the intense curvature at the monopole core, yielding a large $J_Y^A$. As $E_F$ increases, the pocket expands to incorporate regions of progressively weaker curvature, diluting the integral and suppressing the photocurrent. This photocurrent generates a Y-polarized THz field ($E_Y$)[20].

The second contribution is the circular photon drag effect[24,25,30], which arises from the transfer of photon crystal (linear) momentum to the photoexcited carriers. Under oblique incidence at angle $\theta$, the in-plane photon wavevector component $q_x = (\omega/c)\sin\theta$ imparts a net momentum kick during the interband optical transition. The circular polarization state selectively enhances this drag by maximizing the asymmetric occupation of k-states during excitation, producing a net photocurrent along the XZ plane and consequently an X polarized THz ($E_{XZ}$). Importantly, this expression contains no Fermi surface integral and no Berry curvature. The drag current depends only on the photon momentum (set by the incidence geometry), the pump helicity, and the interband absorption at the pump photon energy (~1.55 eV for 800 nm excitation). Since the Fermi level in $Cd_3As_2$ lies only ~210 meV above the Dirac point, electrostatic gating shifts $E_F$ by an amount far too small to modify the optical absorption at the pump frequency through Pauli blocking. The CPDE driven X polarized THz therefore remains gate independent.

The superposition of $E_Y$ and $E_{XZ}$, which are orthogonal and carry an intrinsic phase difference of $\pi/2$ fixed by the excitation geometry (details in supplementary section S4 and S5), produces elliptically polarized THz emission. Since only $E_Y$ responds to the gate voltage while $E_{XZ}$ and the phase difference between them remain constant, sweeping $V_G$ selectively tunes the amplitude ratio and thereby chirality of the emitted THz.

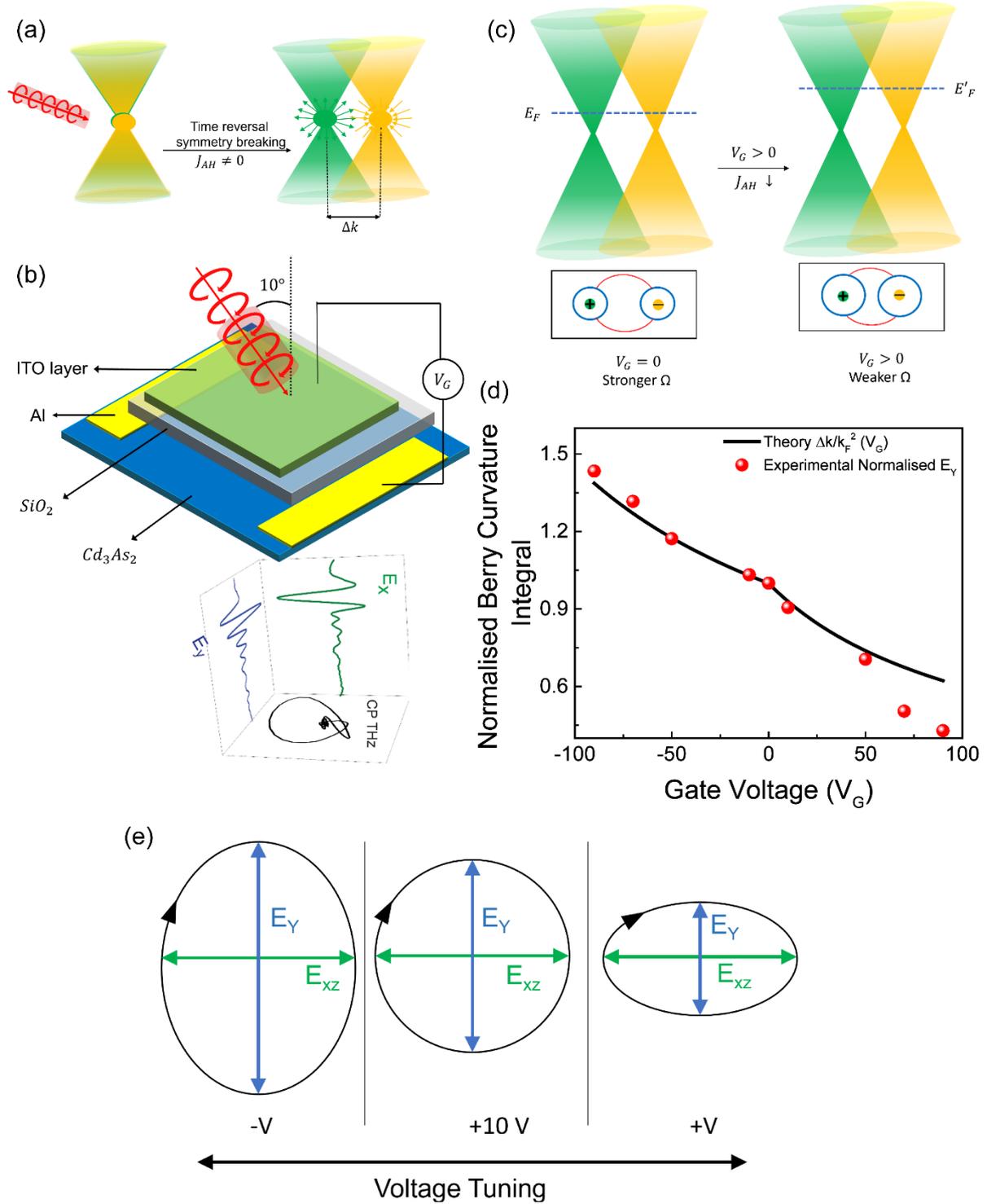

**Figure 1: All-electrical tuning of chiral THz emission from Cd₃As₂;** (a) Floquet band engineering in a 3D Dirac semimetal. Circularly polarized (CP) photoexcitation breaks time-reversal symmetry, splitting the doubly degenerate Dirac node into a pair of Weyl nodes separated by $\Delta k$ in momentum space; (b) Device schematic for electrically tunable chiral THz emission. A 40 nm Cd₃As₂ thin film grown on a c-cut sapphire substrate is gated via an out-of-plane capacitive structure comprising Al electrodes, a 1 $\mu m$ SiO₂ dielectric spacer, and a THz-transparent ITO-coated PET top electrode. CP femtosecond pulses (800 nm, 35 fs) illuminate the sample at 10° oblique incidence, simultaneously generating a Y-polarized THz component (E_Y, blue) from the Floquet anomalous Hall photocurrent and an X polarized THz component (E_{XZ}, green) from the circular photon drag effect (CPDE). The superposition of the two orthogonal, phase-shifted components produces elliptically/circularly polarized THz radiation; (c) Mechanism of gate-tunable Floquet photocurrent. At V_G = 0, the Fermi level sits near the Weyl nodes

where Berry curvature $\Omega$ is strong, yielding a large $E_Y$. Positive $V_G$ shifts $E_F$ away from the nodes, diluting $\Omega$ over the enlarged Fermi pockets and suppressing $E_Y$; (d) Gate voltage dependence of the normalised Berry curvature integral; experimentally measured normalised $E_Y$ (red circles) and the theoretical prediction $\Delta k / k_F^2 (V_G)$ (solid line), showing agreement across the full $\pm 90$ V range. (e) All-electrical chirality control of THz emission from $Cd_3As_2$. $V_G$ modulates only $E_Y$ while $E_{XZ}$ remains fixed. THz polarization evolves from vertically elongated (−V) through circular (+10 V) to horizontally elongated (+V), enabling all-electrical control of THz chirality.

Figure 1(b) shows the gated device structure. A 40 nm $Cd_3As_2$ thin film grown on c-cut sapphire is capped with an out-of-plane capacitive gate stack consisting of Al electrodes, a 1 $\mu m$ $SiO_2$ dielectric spacer, and a THz-transparent ITO-coated PET top electrode. Circularly polarized 800 nm, 35 fs pulses illuminate the device at 10° incidence, simultaneously exciting both THz generation mechanisms. The gate voltage $V_G$ applied across the dielectric tunes the Fermi level within the $Cd_3As_2$ film without modifying the excitation geometry.

Figures 1(c), 1(e) summarize the expected gate response schematically. At $V_G = 0$, $E_F$ sits close to the Weyl nodes where the Berry curvature integral is maximized, yielding a larger $E_Y$ compared to $E_{XZ}$ and a vertically elongated polarization ellipse. Positive $V_G$ shifts $E_F$ upward, enlarging the Fermi pockets and suppressing $E_Y$ while $E_{XZ}$ remains fixed. At +10 V the two amplitudes approximately equalize, and the intrinsic $\pi/2$ phase difference transforms the emission into a circularly polarized THz pulse. Further positive bias continues to suppress $E_Y$, compressing the ellipse toward horizontal linear polarization. Negative $V_G$ produces the complementary evolution, shrinking the Fermi pockets, enhancing $E_Y$, and driving the polarization toward vertical elongation. Figure 1(d) shows the normalised Berry curvature integral as a function of gate voltage. The experimental values, extracted as the normalised peak $E_Y$ amplitude, follow the theoretical $\Delta k / k_F^2 (V_G)$ curve closely across the full $\pm 90$ V range, confirming that Fermi pocket reshaping quantitatively governs the Berry curvature sampled by the photoexcited carriers. The slight deviation of the experimental points below the theoretical curve at large positive gate voltages ($V_G > 50$ V) arises from the breakdown of the fixed screening length assumption in the model. As the injected carrier density increases with positive bias, the Thomas-Fermi screening length self-consistently decreases below the fixed $d_{eff} = 15$ nm assumed in the calculation, further confining the accumulated electrons into a thinner surface layer than the model predicts. This increases the effective three-dimensional carrier density and Fermi wavevector beyond the model estimate, resulting in a stronger dilution of the Berry curvature integral and a larger suppression of $E_Y$ than the linear screening model captures.

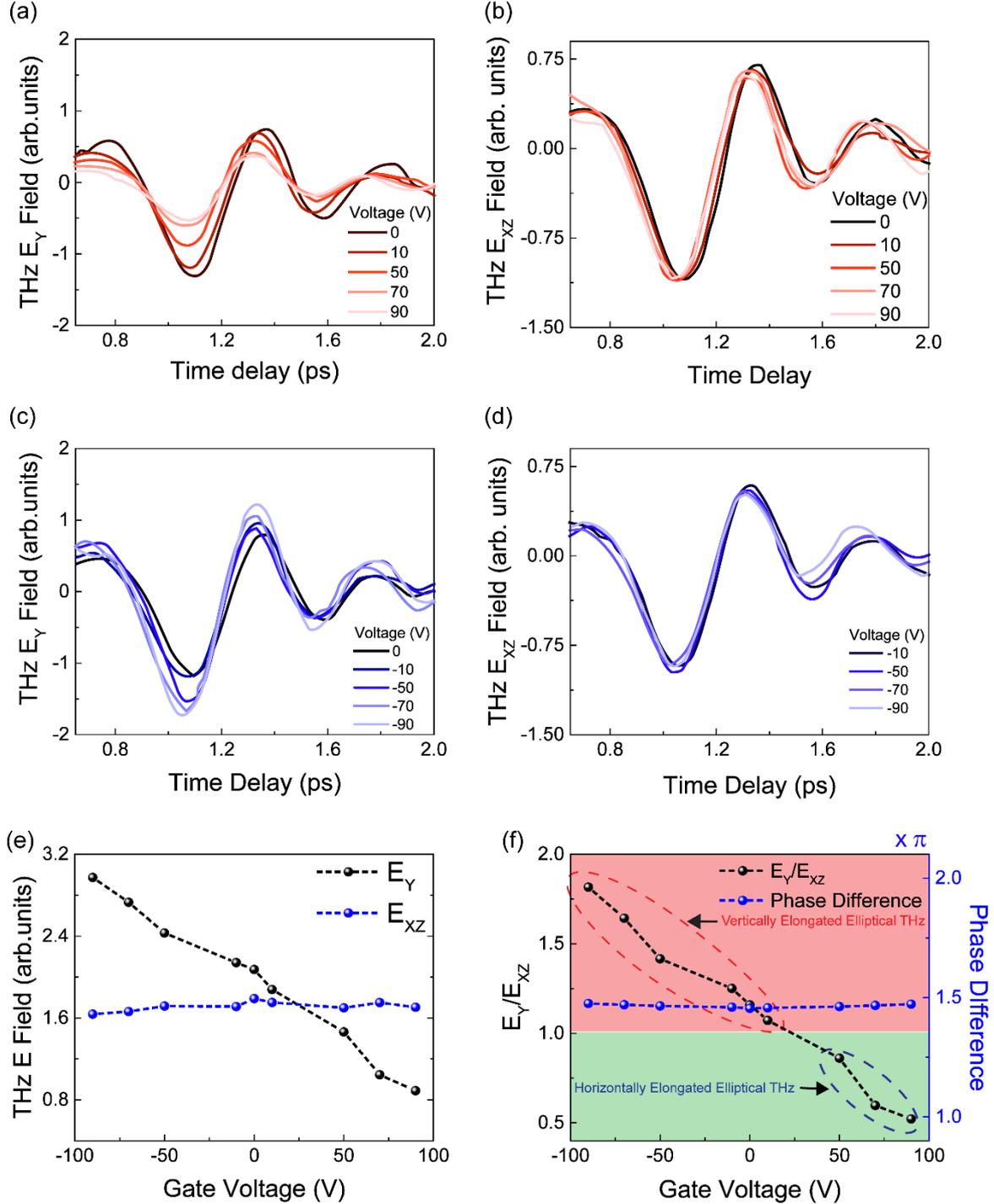

**Figure 2: Selective electrical tuning of the Floquet THz component.** (a) Y polarized THz pulse ($E_Y$) under left circularly polarized pump at 10° incidence for positive gate voltages (0–90 V). $E_Y$ decreases with increasing voltage due to suppression of the Floquet anomalous Hall photocurrent. (b) X polarized THz pulses ($E_{XZ}$) for the same positive gate voltages. $E_{XZ}$ remains unchanged, confirming the voltage independence of the circular photon drag effect. (c) $E_Y$ for negative gate voltages (0 to −90 V). $E_Y$ increases with increasing negative bias as the Fermi level shifts toward the Dirac point, enhancing the Berry curvature flux through due to decrease in Fermi pocket size. (d) $E_{XZ}$ for the same negative gate voltages. $E_{XZ}$ again remains unaffected by the applied bias. (e) Peak-to-peak amplitude of $E_{XZ}$ and $E_Y$ as a function of gate voltage across the full −90 V to +90 V range. $E_Y$ exhibits a strong voltage dependence, whereas $E_{XZ}$ remains approximately constant throughout. (f) Amplitude ratio $E_Y$ / $E_{XZ}$ and phase difference $\delta$ between $E_Y$ and $E_{XZ}$ (blue, right axis) as a function of gate voltage. The ratio crosses unity at ~+10 V, marking the transition from vertically elongated elliptical THz ($E_Y$ / $E_{XZ}$ > 1, pink region)

through circular polarization to horizontally elongated elliptical THz ($E_Y$ / $E_{XZ}$ < 1, green region). The phase difference remains approximately $-\pi/2$ across the entire voltage range, confirming that gating tunes the ellipticity while preserving the handedness of the emitted THz.

Figure 2 presents the experimental realization of this mechanism. Under left circularly polarized excitation, $E_Y$ exhibits a systematic decrease with increasing positive gate voltage from 0 to +90 V as shown in Figure 2(a), consistent with the expected suppression of the Floquet anomalous Hall photocurrent as the Fermi pockets expand away from the Weyl nodes. In contrast, the simultaneously measured $E_{XZ}$ pulses remain unchanged across the same voltage range (Figure 2(b)), confirming the gate insensitivity of the circular photon drag contribution. The gate insensitivity of $E_{XZ}$ is expected as the drag current depends on the photon wavevector component along the surface and the optical absorption coefficient at the pump frequency, neither of which is appreciably modified by the Fermi level shifts accessible via electrostatic gating[24,25,30]. Under negative bias (Figures 2(c), (d)), the complementary behaviour is observed. $E_Y$ increases progressively from 0 to $-90$ V as the Fermi level shifts toward the Dirac point, shrinking the Fermi pockets and enhancing the sampled Berry curvature flux. $E_{XZ}$ again remains unaffected. Figure 2(e) summarizes the peak-to-peak amplitudes of both components across the full $-90$ V to +90 V range, making the selective tunability visually explicit. $E_Y$ varies strongly and monotonically with $V_G$ while $E_{XZ}$ remains approximately constant throughout. The total modulation of $E_Y$ in positive voltage window is 60% while in negative voltage window, the modulation is 49%.

The amplitude ratio $E_Y$/$E_{XZ}$ and the intercomponent phase difference $\delta$ are plotted as functions of gate voltage in Figure 2(f). The ratio decreases monotonically from values well above unity at negative bias to well below unity at positive bias, crossing unity at approximately +10 V. Crucially, the phase difference $\delta$ remains locked near $\pi/2$ across the entire voltage range. This invariance confirms that gating modulates only the amplitude balance between the two orthogonal THz fields without perturbing their temporal relationship. The combination of a continuously tunable amplitude ratio and a fixed phase offset is precisely the condition required for deterministic traversal of the THz polarization state across the Poincaré sphere via a single electrical control parameter.

Figure 3 maps the voltage-dependent polarization state by plotting $E_Y(t)$ versus $E_{XZ}(t)$ parametrically at representative gate voltages. At 0 gate voltage (Figure 3(a)), the THz field traces a slightly vertically elongated ellipse, reflecting the larger $E_Y$ amplitude at zero bias. At +10 V (Figure 3(b)), the two amplitudes equalize and the ellipse opens into a nearly circular trace, consistent with the amplitude ratio crossing unity while the $\pi/2$ phase offset is maintained. Further increasing $V_G$ to +50, +70, and +90 V (Figures 3(c–e)) progressively suppresses $E_Y$, compressing the ellipse along the vertical axis until the polarization approaches linear along the horizontal direction. Negative gate voltages produce the opposite trend. At −10 V (Figure 3(f)), $E_Y$ is mildly enhanced, yielding a slightly more vertically elongated ellipse compared to 0 V. By −50 V (Figure 3(g)) and −90 V (Figure 3(h)), the vertical elongation becomes dominant as the enhanced Berry curvature flux drives an increasingly large $E_Y$. Throughout the entire voltage sweep, the rotation direction of the polarization vector remains fixed, confirming that the handedness is preserved while the ellipticity is tuned electrically. This panel-by-panel evolution constitutes a direct, real-space visualization of how Fermi pocket reshaping translates into macroscopic polarization control.

To quantify the electrically driven polarization evolution, Figure 4 presents the extracted Stokes parameters and ellipticity as functions of gate voltage. Figure 4(a) shows the ellipticity angle $\chi$, which reaches a maximum of approximately −42° near +10 V which is close to the $45°$ limit of perfect circular polarization and decreases symmetrically for both positive and negative voltage deviations as the amplitude balance shifts away from unity. The normalized Stokes parameter $S_1$ (Figure 4(b)) transitions from negative values at negative bias, corresponding to vertically dominated polarization, through zero at approximately +10 V, to positive values at large positive bias where horizontal polarization prevails. The zero crossing of $S_1$ coincides with the ellipticity minimum, independently confirming that +10 V is the operating point of near-perfect circular THz emission.

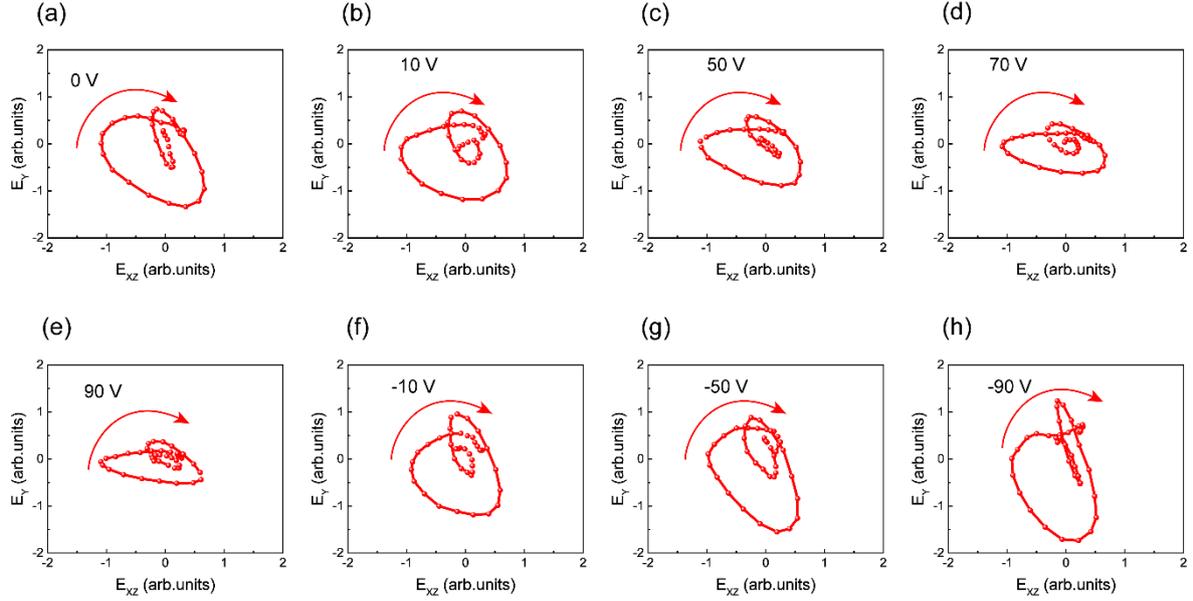

**Figure 3: Voltage-dependent polarization ellipses of the emitted THz field.** (a) Parametric plot of $E_Y(t)$ versus $E_{XZ}(t)$ at 0 V under left circularly polarized pump, showing a slightly vertically elongated ellipse ($E_Y > E_{XZ}$). (b) Polarization ellipse at +10 V, where $E_Y \approx E_{XZ}$ and the THz field is closest to circular. (c) Polarization ellipse at +50 V, with reduced $E_Y$ and a clearly horizontally elongated ellipse. (d) Polarization ellipse at +70 V, where further suppression of $E_Y$ drives the polarization toward linear along $E_{XZ}$. (e) Polarization ellipse at +90 V, showing a strongly $E_{XZ}$ dominated, highly horizontally elongated elliptical polarization. (f) Polarization ellipse at −10 V, where enhancement of the Floquet photocurrent increases $E_Y$ and produces a mildly vertically elongated ellipse. (g) Polarization ellipse at −50 V, exhibiting a more strongly vertically elongated ellipse as $E_Y$ becomes dominant. (h) Polarization ellipse at −90 V, showing a highly vertically elongated state corresponding to maximal enhancement of $E_Y$. Red arrows in all panels indicate the rotation direction of the polarization vector, confirming fixed handedness while the ellipticity is tuned electrically.

The gate-driven evolution of the THz polarization state constitutes a direct mapping of quantum geometric tuning onto a macroscopic optical observable. Figure 4(c) maps the measured polarization states onto the Poincaré sphere where each point corresponds to a unique polarization state[20,31]. The sphere is parametrized by the azimuthal angle $2\psi$ along the equator and the polar angle $2\chi$ along the meridian, with the Cartesian coordinates given by $S_1 = \cos 2\chi \cos 2\psi$, $S_2 = \cos 2\chi \sin 2\psi$, and $S_3 = \sin 2\chi$. The equator represents all linear polarization states, while the north and south poles correspond to left and right circular polarization, respectively. The vertical coordinate $S_3 = \sin 2\chi$ thus directly quantifies the degree of circularity, with $S_3 = -1$ for LCP and $S_3 = +1$ for RCP. Points at intermediate latitudes represent elliptical states of varying eccentricity. The detailed construction of the Poincaré sphere from the measured THz waveforms is described in the Methods section. The data points trace a continuous arc near the LCP pole, demonstrating that the gate voltage drives a smooth trajectory across the sphere rather than discrete switching between a limited set of states. The polarization states of the emitted pulse are tuned from left circularly polarized ($\chi = -42°$) when gate voltage is 10 V, to vertically elongated elliptical polarization ($\chi = -27°$) when gate voltage is $-90$ V, and to vertically elongated elliptical polarization ($\chi = -25°$) when gate voltage is 90 V indicating a continuous electrical control of THz polarization and chirality.

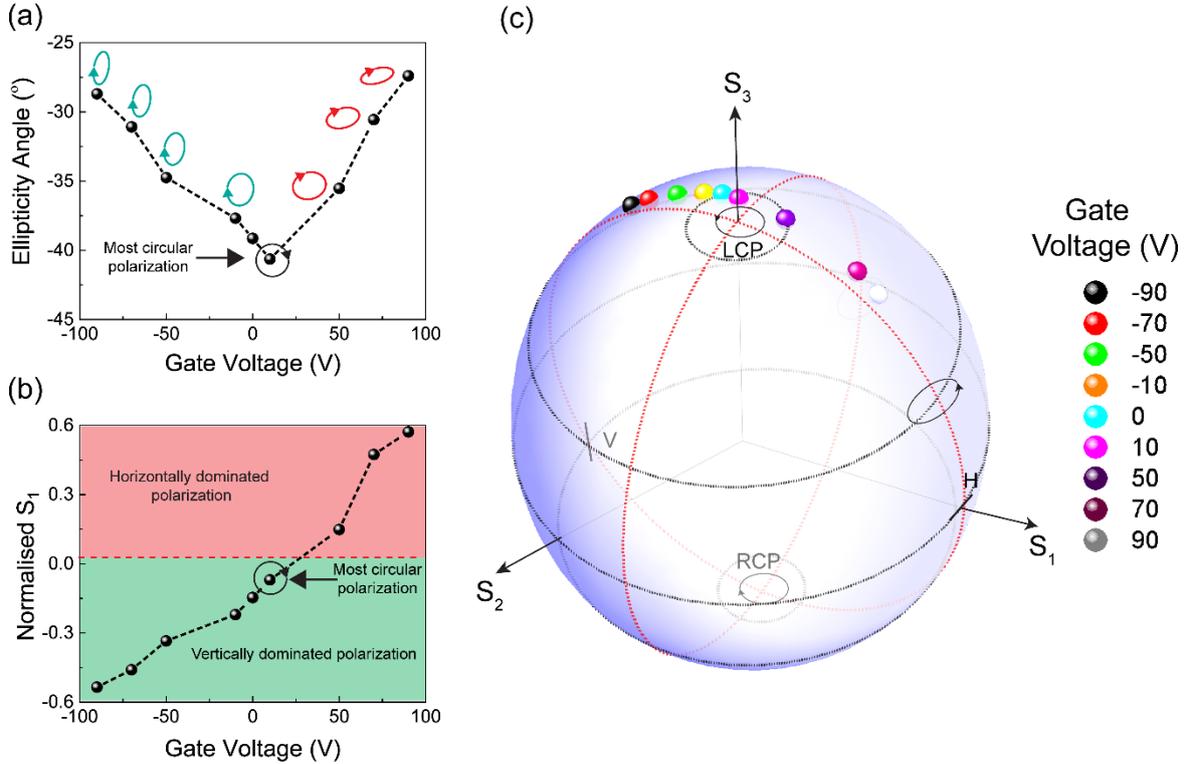

**Figure 4: Quantitative polarization analysis of electrically tuned chiral THz.** (a) Ellipticity angle $\chi$ as a function of gate voltage, with maximum circularity ($\chi \approx -42°$, dashed circle) near +10 V where $E_Y \approx E_{XZ}$. Both positive and negative voltage deviations reduce the circularity as the amplitude balance shifts. (b) Normalized Stokes parameter $S_1$ as a function of gate voltage, transitioning from vertically dominated polarization (Normalised $S_1<0$, green region) through balanced (Normalised $S_1=0$) to horizontally dominated polarization (Normalised $S_1>0$, pink region). The zero crossing at ~+10 V is consistent with the ellipticity minimum in (a). (c) Measured polarization states mapped onto the Poincaré sphere with LCP at the north pole. The data points trace an arc near the LCP pole, demonstrating continuous electrical tuning of the THz polarization state. The sphere is displayed with LCP at the north pole for visual clarity; the data cluster near $\chi \approx -42°$, corresponding to $S_3 \approx -0.995$.

In summary, we have demonstrated that the Fermi pocket size serves as a quantitative tuning parameter for the Berry curvature component of quantum geometry in a topological semimetal, with the continuous gate-driven traversal of the THz Poincaré sphere as the macroscopic optical signature. The same strategy is directly applicable to the nonlinear anomalous Hall effect and bulk photovoltaic effect in other topological semimetals. The room temperature operation, reversibility, and compatibility with standard thin film gating architectures make this approach readily transferable to on-chip geometries, pointing toward electrically programmable quantum-geometric control of chiral THz sources for spectroscopy, imaging, and communications

**Methods**

***Sample preparation***: A 40 nm $Cd_3As_2$ thin film was grown by molecular beam epitaxy (PerkinElmer 425B) on a c-cut sapphire substrate. (thickness 420 $\mu m$, $\langle 0001 \rangle$ orientation). The substrate was annealed at 550°C for 30 min to desorb surface contaminants. A 20 nm CdTe buffer layer was first deposited to accommodate the lattice mismatch between $Cd_3As_2$ and sapphire. High-purity (99.999%) Cd and As source materials were then co-evaporated from dual-filament and valve-cracker effusion cells onto the buffer layer at a substrate temperature of 120 °C. Film thickness was monitored *in situ* by reflection high-energy electron diffraction (RHEED).

***THz emission spectroscopy:*** Broadband THz emission was measured by electro-optic sampling in a 1 mm thick $\langle 110 \rangle$-oriented ZnTe crystal. The excitation source was an amplified Ti: sapphire laser (800 nm, 1.55 eV, 35 fs, 1 kHz repetition rate). A beam splitter divided the output into a high-intensity pump arm directed onto the $Cd_3As_2$ film and a low-intensity probe arm used for detection. A quarter-wave plate in the pump arm controlled the polarization state of the excitation beam. The emitted THz pulse was collected and focused onto the ZnTe detector by a set of four off-axis parabolic mirrors, and a mechanical delay stage provided temporal overlap between the THz and probe pulses at the detector crystal. The THz-induced birefringence in ZnTe was read out by passing the co-propagating probe beam through a quarter-wave plate and Wollaston prism and measuring the differential intensity of the resulting s- and p-polarized components with a balanced photodiode. The photodiode signal was pre-amplified and fed into a lock-in amplifier referenced to an optical chopper to improve the signal-to-noise ratio. To resolve the orthogonal THz field components $E_Y$ and $E_{XZ}$ independently, two wire-grid polarizers (WGP) were inserted into the THz beam path (Supplementary Section S1). WGP-1 selected either the $E_Y$ or $E_{XZ}$ component, while WGP-2 was fixed at 45° to rotate the transmitted polarization into a common orientation for uniform detection at the ZnTe crystal. The entire beam path was enclosed in a dry nitrogen atmosphere to suppress water vapour absorption. A detailed schematic of the setup is provided in Supplementary Section S1.

***Construction of Poincaré sphere***: To have a better visualization of the polarization state of the emitted THz pulse a Poincaré sphere was constructed[31], as shown in Figure 4(c). In a unit sphere, the cartesian coordinates and the spherical coordinates are related via the following equations:

$$x = \cos(2\chi)\cos(2\psi), \quad 0 \le \psi < \pi$$

$$y = \cos(2\chi)\sin(2\psi), \quad -\frac{\pi}{4} < \chi < \frac{\pi}{4}$$

$$z = \sin(2\chi)$$

where, $\psi$ and $\chi$ are the spherical coordinates and $x^2 + y^2 + z^2 = 1$ is the equation of the unit sphere. The spherical coordinates indicated by $\psi$ and $\chi$ of the Poincaré sphere are defined as the parameters of the polarization ellipse (azimuth angle: $\psi$ and ellipticity angle: $\chi$) from the following equations.

$$\tan 2\psi = \frac{2E_Y E_{XZ}}{E_{XZ}^2 - E_Y^2}\cos\delta \qquad 0 \le \psi \le \pi$$

$$\sin 2\chi = \frac{2E_{XZ} E_Y}{E_{XZ}^2 + E_Y^2}\sin\delta \qquad -\frac{\pi}{4} \le \chi \le \frac{\pi}{4}$$

where, $\delta$ is the phase difference between the X and Y polarized light,

$E_{XZ}$ is the electric field amplitude of X-polarized light

$E_Y$ is the electric field amplitude of Y-polarized light

The phase difference $\delta$ for all the pump polarizations are calculated from the Fast Fourier Transform (FFT) of the emitted THz time pulses, as shown in main text Figure 2(f). The electric field amplitude for X and Y polarized light $E_{XZ}$ and $E_Y$ are recorded separately by employing a combination of 2 wire-grid polarizers, as shown in Supplementary section S1.

***Berry curvature integral calculation***: The normalised Berry curvature integral was calculated as $\Delta k / k_F^2(V_G)$, where $k_F(V_G) = E_F(V_G)/\hbar v_F$ is the gate-dependent Fermi wavevector, $\Delta k$ is the Weyl node separation fixed by the photoexcitation, and $v_F = 1.5 \times 10^6$ m/s is the Fermi velocity of Cd$_3$As$_2$. The gate-induced Fermi energy $E_F(V_G)$ was determined from the carrier density $n(V_G) = C_{ox}V_G/e$, using an asymmetric screening model with a depletion depth of 40 nm (negative $V_G$) and an accumulation depth of 15 nm (positive $V_G$) to account for the asymmetric charge redistribution across the 40 nm film. Full derivation is provided in Supplementary Section S7.

## Acknowledgments

S.M. thanks Dr. Manoj Gupta and Dr. Baolong Zhang for valuable discussions and suggestions. S.M, T.C.T, and R.S acknowledge funding support from National Research Foundation (NRF) Singapore, Grant no. NRF-MSG-2023-0002 (NCAIP).

F.X. was supported by the National Natural Science Foundation of China (52225207) and the Shanghai Pilot Program for Basic Research - FuDan University 21TQ1400100 (21TQ006).

## Author Contributions

S.M, R.S conceived the project. S.M and R.S designed the THz emission and electrical tuning experiments; F.X prepared the $Cd_3As_2$ thin film samples; S.M performed all the THz emission measurements and experimental analysis with the help from T.C.T; S.M and R.S analyzed and discussed the results; S.M. and R.S. wrote the manuscript with input from all the authors; F.X and R.S led the overall project.

## Competing Interests

The authors declare no competing interests

# Supplementary Information

## Section S1: Terahertz emission spectroscopy setup

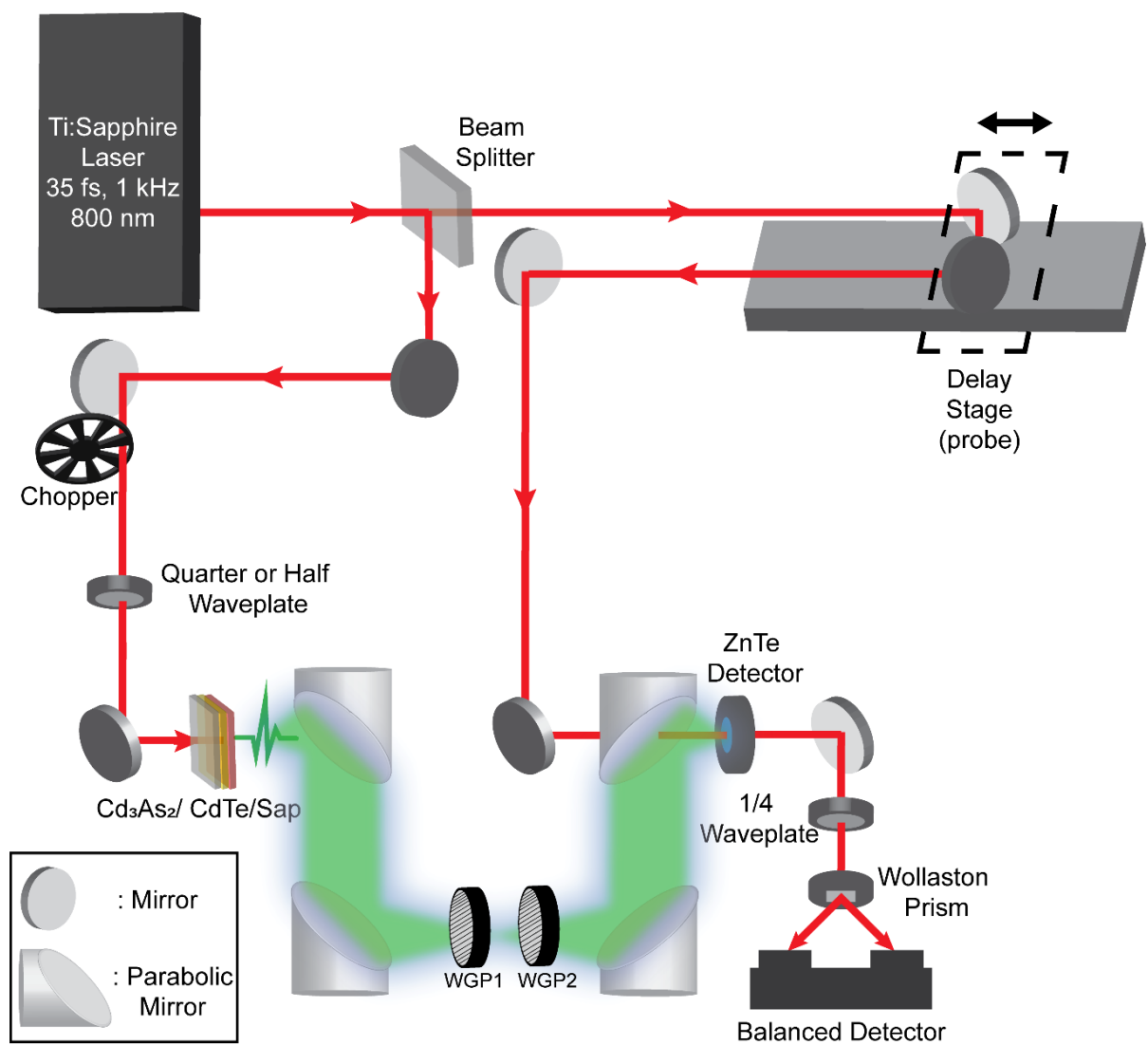

**Figure S1: THz emission spectroscopy set up for chiral THz generation from $Cd_3As_2$**

Figure S1 shows the Chiral THz emission spectroscopy setup with Cd₃As₂ as a Terahertz emitter. A quarter waveplate (QWP) is introduced in the pump path to photoexcite Cd₃As₂ with different polarized light. 2 wire grid polarizer was employed before the detector to separate the $E_Y$ and $E_{XZ}$ component of the emitted THz. The 2nd QWP is always kept at 45° and the 1st QWP is kept at 0°/90° to extract $E_Y$ and $E_{XZ}$ respectively.

**Section S2: Quantitative estimation of Fermi level shift with applied voltage Induced carrier density:**

The capacitive device has measured capacitance $C = 1.15$ nF over an overlap area of 0.64 cm². At the maximum bias $V = 90$ V, the induced surface charge is:

$$Q = CV = 103.5 \text{ nC},$$

$$\Delta n_{2D} = \frac{Q}{eA} \approx 1.0 \times 10^{12} \text{ cm}^{-2}$$

**Fermi level shift:** For a 3D Dirac semimetal with $E_F = 210$ meV and Fermi velocity $v_F = 1.5 \times 10^6$ m/s, the bulk carrier density is

$$n_0 = N_D k_F^3/(3\pi^2) \approx 6.5 \times 10^{17} \text{ cm}^{-3}$$

where $N_D = 2$ is the number of Dirac nodes and $k_F = E_F/(\hbar v_F) \approx 2.1 \times 10^8$ m$^{-1}$.

The Thomas-Fermi screening length in Cd3As2 is

$$\lambda_{TF} = \sqrt{\varepsilon_0 \varepsilon_r/(e^2 g(E_F))} \approx 21 \text{ nm}$$

Here $\varepsilon_r \approx 36$[1]

$$g(E_F) = N_D E_F^2/(\pi^2 \hbar^3 v_F^3).$$

Since $\lambda_{TF} < d = 40$ nm, the field partially screens within the film, and the induced carriers accumulate primarily in a surface layer of thickness $\sim \lambda_{TF}$.

$$\Delta n_{3D} = 1.01 \times 10^{16}/(21 \times 10^{-9}) = 4.87 \times 10^{17} \text{ cm}^{-3}$$

Since $n \propto E_F^3$ for 3D Dirac dispersion, the Fermi level shift is:

$$\Delta E_F = E_F \left[ \left( 1 + \frac{\Delta n}{n_0} \right)^{1/3} - 1 \right] \approx 43 \text{ meV}$$

corresponding to $E_F$ shifting from 210 meV to approximately 250 meV.

**Effect on anomalous Hall photocurrent:** The Berry curvature near each Floquet-Weyl node follows the monopole form $\Omega \sim 1/k^2$. The effective anomalous Hall conductivity sampled by carriers at the Fermi surface scales approximately as $\sigma_{xy,\text{eff}} \propto \Delta k/k_F^2$, where $\Delta k = e^2 v_F E_{\text{pump}}^2/(\hbar^2 \omega^3)$ is set by the optical drive and is independent of gating. As $E_F$ (and hence $k_F$) increases, the Berry curvature is increasingly diluted over the enlarged Fermi surface, reducing the photocurrent.

Since the Berry curvature near each Floquet-Weyl node follows the monopole form $\Omega(\mathbf{k}) = \frac{C}{2k^2}$, the Berry curvature integral over the Fermi pocket scales as $\Delta k / k_F^2$, where $\Delta k$ is the Weyl node separation. Gate driven Fermi pocket reshaping therefore provides direct electrical control over the quantum geometric contribution to the photocurrent, as this Berry curvature integral which is the key quantum geometric invariant governing the anomalous Hall response scales inversely with $k_F^2(V_G)$.

## Section S3: Disentanglement of different contributions and reconstructed THz signal due to helicity dependent, polarization dependent and photothermal term

The peak-to-peak amplitude of the emitted signal was measured as a function of pump polarization, which was varied by adjusting the Quarter Wave Plate (QWP) angle. The polarization dependence of the amplitude of the emitted THz pulse, displayed in Figure S2(a), was fitted to the phenomenological model described below in equation S1[2–4]. This fitting enabled the extraction of coefficients representing the distinct contributions to THz emission from $Cd_3As_2$, as illustrated in Figure S2(b)

$$E_{THz} = Csin2\alpha + L_1sin4\alpha + L_2cos4\alpha + D \tag{S1}$$

Here the first term ($Csin2\alpha$) explains the helicity and polarization dependent contribution towards THz emission, the second ($L_1sin4\alpha$) and third ($L_2cos4\alpha$) terms capture the polarization dependent but helicity independent THz emission, and the last term ($D$) describes the helicity and polarization independent THz emission due to ultrafast photothermal effect[5]. $L_1$ describes the shift currents which arise due to a spatial shift in the charge carrier position resulting in an interband optical excitation generating a time-dependent dipole moment. Since these currents are symmetry-constrained, they require an electric field component normal to the surface. Therefore, at normal incident angle, their contribution towards THz emission is negligible as shown in Figure S2(b). Moreover, the shift current contribution vanishes at both linear and circular polarized photoexcitation. $L_2$ and $D$ depict the thermal effects associated with linear absorption with $L_2$ describing the polarization dependent modulation of photothermal effect and $D$ accounting for the polarization and helicity independent ultrafast photothermal effect. Quantitatively, the helicity dependent term is about twice the helicity independent term and is of negative value as shown in Figure 2(b) indicating that the THz emission is predominantly from helicity dependent mechanism when time reversal symmetry is broken. Additionally, it was also observed that $L_2$ and $D$ have comparable values.

To gain further understanding of the system, the contributions of the THz pulse from individual components described in Equation S1 were reconstructed and presented for both linear and circular polarized pumps in Figure S2(c-f). For these polarizations, THz pulse contribution from the second term ($L_1 sin4\alpha$) will be zero since $sin4\alpha$ equals zero when $\alpha$ = 0°, 45°, 90°, or 135°. Under linear pump polarization ($\alpha$ = 0°, 90°), the helicity-dependent term ($C sin2\alpha$) also vanishes, but the remaining two terms ($L_2 cos4\alpha$ and $D$) add constructively to produce a stronger THz pulse, as illustrated in Figures S2(c) and S2(e). When the pump is left circularly polarized in nature ($\alpha$ = 45°), the helicity-dependent term ($C sin2\alpha$) becomes negative due to the negative value of $C$, resulting in phase reversal in the emitted THz pulse, as shown in Figure S2(d). Notably, the other two terms ($L_2 cos4\alpha$ and $D$) cancel each other out, as $L_2$ and $D$ are of comparable magnitude (see Figure S2(b)), making the resultant THz pulse entirely attributed to the helicity- dependent term. When the pump polarization is right-circular ($\alpha$ = 135°), the phase of the THz pulse is opposite to the emitted THz when the pump is left circular ($\alpha$ = 45°) in nature, and the helicity-dependent component of the THz pulse is similarly inverted, as depicted in Figure S2(d) and S2(f). Hence the at circularly polarized pump conditions all the other contributions are completely cancelled out.

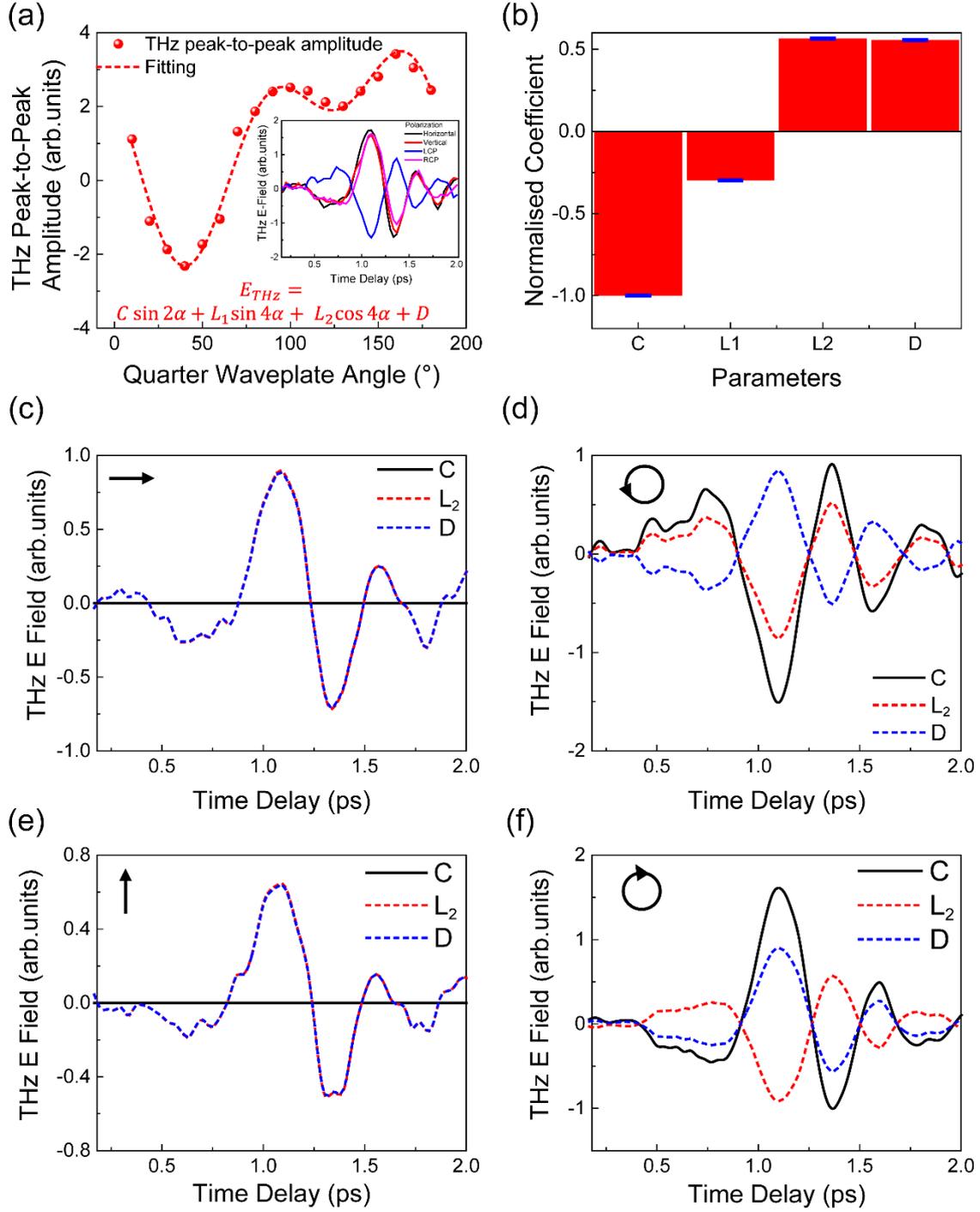

**Figure S2: Disentanglement of different contributions and reconstructed THz signal due to helicity dependent, polarization dependent and photothermal term** (a) Extracted THz peak-to-peak amplitude at different QWP angle ($\alpha$) for optical pump light. Dotted line is the fitted curve according to the phenomenological equation 2, inset shows the THz pulses at both the linear and circularly polarized pumps; (b) Normalized extracted fitting parameters attributed to the factors affecting the emitted THz; Reconstructed THz pulse due to each factor when pump polarization is (c) Horizontal (d) Left circular (e) Vertical (f) Right circular. When the pump beam is circularly polarized, $L_2$ and $D$ cancel each other resulting in THz emission mainly due to helicity dependent term (Floquet band engineering)

**Section S4: Evidence of presence of CPDE at oblique incidence**

It is necessary to show an experimental proof of presence of circular photon drag effect (CPDE) in Cd₃As₂ when pumped with oblique incidence. Following experimental evidence can prove the generation of circular photon drag current

1. As we know CPDE is a resultant of angular momentum transfer between photons and electrons. As a result, the direction of photocurrent will reverse if the momentum of the incident light is reversed. To show that, at oblique incidence, longitudinal THz is a result of CPDE, we changed the angle of incidence from -20° to 20°. As shown in Figure S3(a) the phase of the emitted THz is reversed when the incident angle is changed from -20° to 20° and the amplitude is almost 0 when the pump is normally incident. However, the direction of the transverse component, which is a resultant of Floquet band engineering, is independent of the incident angle as shown in Figure S3(b). The slight change in the amplitude of the photocurrent can be a consequence of possible presence of out of plane spin texture of the Dirac cone at oblique incident.

2. Secondly, under normal incidence, the longitudinal component of the THz field is effectively suppressed, consistent with the absence of CPDE in this configuration.

(a) (b)

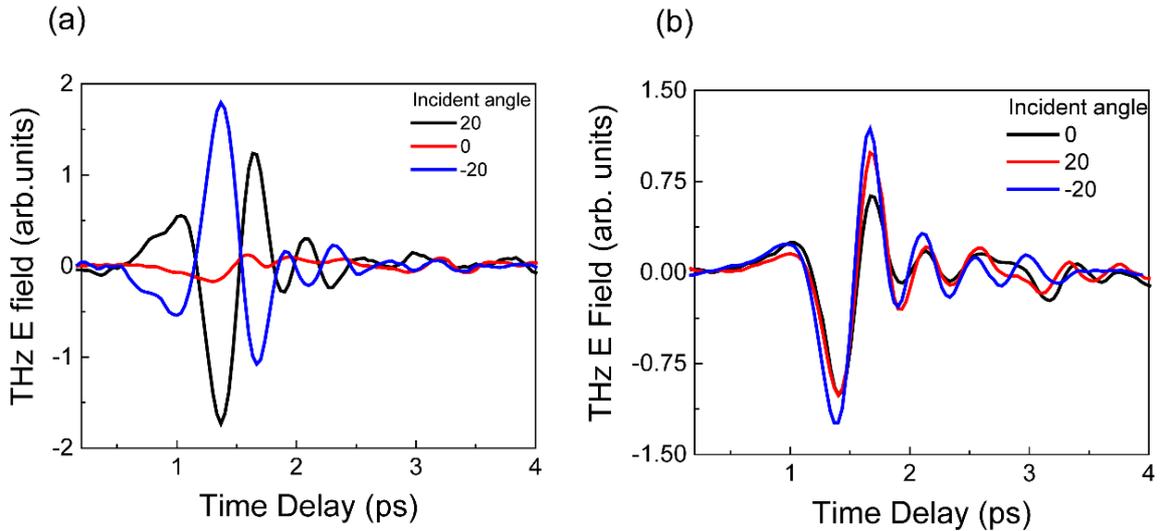

**Figure S3: Incident angle dependence of longitudinal and transverse component of the emitted THz** (a) Longitudinal component of the emitted THz. There is a phase flip when the incident angle is reversed from -20° to 20° and negligible THz at normal incidence indicating that circular photon drag effect is the dominating mechanism for longitudinal component. (b) Y polarized THz at different incident angle. The phase of the emitted pulse is independent of the direction of the photoexcitation, indicating the absence of circular photondrag effect for X polarized THz.

**Section S5: Excitation geometry dependence of phase of $E_{xz}$**

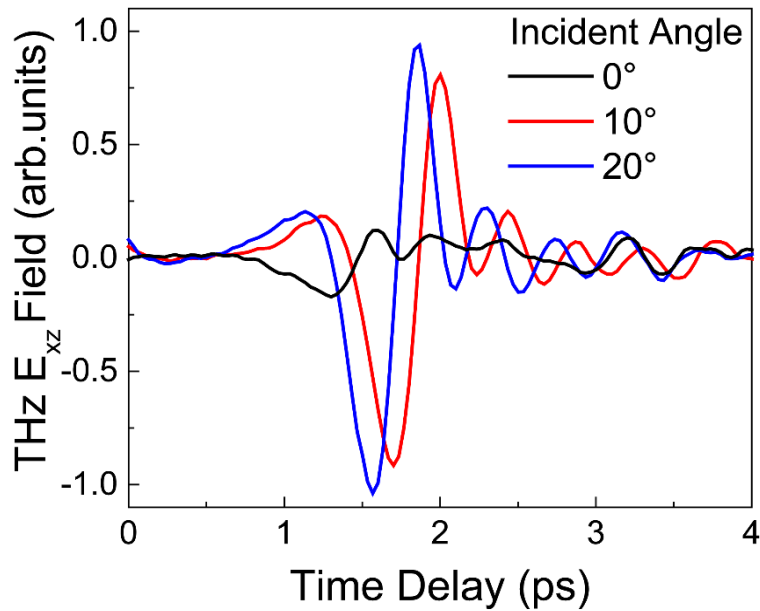

**Figure S4: Manipulation of phase of the X component by excitation geometry**

Figure S4 indicates, when illuminated by circularly polarized light at normal incidence, the X polarized component of the emitted THz pulse is almost negligible and it increases as the incident angle is increased, signifying a pronounced angular dependence. Additionally, there is a shift in the THz pulse indicative of change in phase of the emitted pulse.

**Section S6: Voltage dependent stokes parameters and degree of polarization**

To provide a complete characterization of the gate-tunable THz polarization state, Figure S5 presents the remaining Stokes parameters $S_3$ and $S_2$, along with the degree of circular and linear polarization, as functions of gate voltage. The near unity degree of circular polarization and near zero degree of linear polarization at +10 V independently confirm the ellipticity analysis presented in the main text. The invariance of $S_2 \approx 0$ across all voltages confirms that the polarization axis does not rotate with gating and only the ellipticity changes, consistent with the physical picture in which the gate modulates the $E_Y$ amplitude while leaving both the $E_{xz}$ amplitude and the phase difference between them remains unchanged.

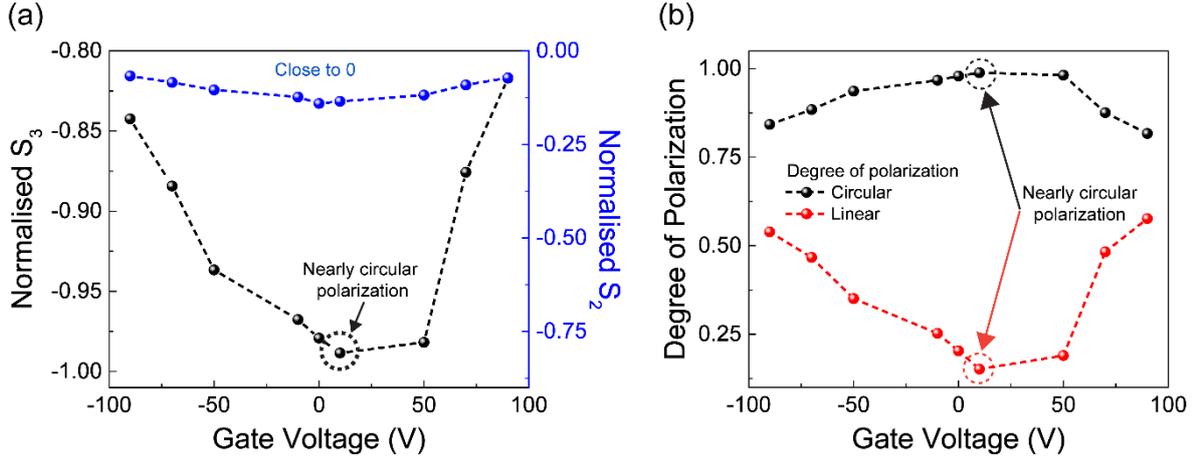

**Figure S5: Stokes parameter analysis and degree of polarization as a function of gate voltage** (a) Normalized Stokes parameters $S_3$ (black, left axis) and $S_2$ (blue, right axis) as a function of applied gate voltage from −90 V to +90 V. $S_3$ reaches a minimum near −1 at approximately +10 V, indicating maximum left circular polarization, and relaxes toward less negative values at both positive and negative voltage extremes as the ellipticity decreases. $S_2$ remains close to zero throughout the entire voltage range, confirming that the polarization evolution is confined to the $S_1$–$S_3$ plane of the Poincaré sphere with no azimuthal rotation of the ellipse. (b) Degree of circular polarization (black) and degree of linear polarization (red) as a function of gate voltage. The degree of circular polarization reaches a maximum (~1.0) and the degree of linear polarization reaches a minimum (~0.1) near +10 V, consistent with the near-perfect circularly polarized THz emission at this operating point. The complementary evolution of the two quantities confirms that the gate voltage drives a clean transition between elliptical and circular states without introducing spurious polarization mixing.

## Section S7: Berry curvature integral calculation

The Berry curvature is the imaginary part of the quantum geometric tensor $Q_n(\mathbf{k}) = g_n(\mathbf{k}) + \frac{i}{2}\Omega_n(\mathbf{k})$, where $g_n(\mathbf{k})$ is the Fubini-Study quantum metric of band $n$ and $\Omega_n(\mathbf{k})$ is the Berry curvature. In the context of the anomalous Hall photocurrent, only the antisymmetric Berry curvature component contributes to the transverse response, making the THz emission a direct probe of this quantum geometric quantity.

In a Weyl semimetal, each nodal point carries a topological monopole charge $C = \pm 1$ and acts as a quantised source of Berry curvature in momentum space. Near the Weyl node, the Berry curvature takes the form of a magnetic monopole field:

$$\Omega(k) = \frac{C}{2k^2}$$

where $k = |\mathbf{k}|$ is the distance from the nodal point in momentum space. This $1/_{k^2}$ divergence concentrates the largest Berry curvature flux at the Weyl node core. The anomalous Hall photocurrent $J_Y$ is proportional to the Berry curvature integrated over the occupied states enclosed within the Fermi pocket:

$$J_Y \propto \int_{\text{pocket}} \Omega(k) \; d^2k \qquad \text{(S2)}$$

For a 2D isotropic Fermi pocket of radius $k_F$ enclosing a single Weyl node, converting to polar coordinates gives:

$$\int_0^{k_F} \Omega(k) \cdot 2\pi k \; dk = \pi C \int_0^{k_F} \frac{dk}{k} = \pi C \ln\left(\frac{k_F}{k_0}\right) \qquad \text{(S3)}$$

where $k_0$ is a short-distance cutoff set by the lattice scale. For the pair of Weyl nodes separated by $\Delta k$ in momentum space, Taylor expanding around the equilibrium Fermi wavevector $k_{F^0}$, the normalised gate-dependent Berry curvature integral is:

$$\Omega_{\text{norm}}(V_G) = \frac{\Delta k}{k_F^2(V_G)} \qquad \text{(S4)}$$

## Gate voltage to carrier density

The electrostatic gate modifies the carrier density in the Cd$_3$As$_2$ thin film via a parallel-plate capacitor model. The oxide capacitance per unit area is:

$$C_{\text{ox}} = \frac{\varepsilon_0 \, \varepsilon_r}{d_{\text{ox}}} \qquad \text{(S5)}$$

where $\varepsilon_0$ = 8.854 × 10$^{-12}$ F/m, $\varepsilon_r$ = 3.9 (SiO$_2$), and $d_{\text{ox}}$ = 1 $\mu$m, giving $C_{\text{ox}}$ = 3.45 × 10$^{-8}$ F/m². Note that the theoretical plate capacitance used here is consistent with the measured device capacitance of 1.15 nF over an active area of 0.64 cm², giving a measured C/A = 1.80 × 10$^{-8}$ F/m², within a factor of 2 of the theoretical value. The small discrepancy reflects fringing fields and interfacial effects.

The gate-induced change in areal carrier density is:

$$\Delta n_{\text{2D}}(V_G) = \frac{C_{\text{ox}} \, V_G}{e} \qquad \text{(S6)}$$

where $e$ = 1.602 × 10$^{-19}$ C. The three-dimensional carrier density is obtained by dividing by the effective electrical thickness $d_{\text{eff}}$ of the film (see Section below):

$$\Delta n_{\text{3D}}(V_G) = \frac{\Delta n_{\text{2D}}(V_G)}{d_{\text{eff}}} \qquad \text{(S7)}$$

**Carrier density to Fermi energy**

$Cd_3As_2$ is a 3D Dirac semimetal with linear isotropic dispersion $E = \hbar v_F k$. The equilibrium Fermi energy at $V_G = 0$ is $E_{F^0} = 210\ meV$. The equilibrium Fermi wavevector is:

$$k_F^0 = \frac{E_F^0}{\hbar v_F} \qquad (S8)$$

with $v_F = 1.5 \times 10^6$ m/s. For a 3D Dirac semimetal, carrier density and Fermi wavevector are related by:

$$n_0 = \frac{(k_F^0)^3}{3\pi^2} \qquad (S9)$$

Under an applied gate voltage, the total carrier density becomes $n(V\_G) = n_0 + \Delta n_{3D}(V_G)$, and the gate-dependent Fermi wavevector is:

$$k_F(V_G) = (3\pi^2\, n(V_G))^{1/3} \qquad (S10)$$

The corresponding gate-dependent Fermi energy is:

$$E_F(V_G) = \hbar v_F\, k_F(V_G) \qquad (S11)$$

**Asymmetric screening model**

The effective depth $d_{eff}$ over which gate induced carriers are redistributed is asymmetric between accumulation and depletion regimes, reflecting the distinct charge redistribution mechanisms in a 40 nm thin film.

Under **depletion** (negative $V_G$), electrons are removed and the depletion front extends across the full film thickness:

$$d_{\text{eff}}^- = d_{\text{film}} = 40\ \text{nm}(V_G < 0) \qquad (S12)$$

Under **accumulation** (positive $V_G$), injected electrons are confined near the gate interface within the Thomas-Fermi screening length $\lambda_{TF}$. For $Cd_3As_2$ at $E_F = 210\ meV$, $\lambda_{TF} \approx 15 - 21$ nm. We use:

$$d_{\text{eff}}^+ = \lambda_{\text{acc}} = 15\ \text{nm}(V_G > 0) \qquad (S13)$$

This asymmetry causes the carrier density change per unit voltage to be larger under accumulation than under depletion, consistent with the steeper suppression of $E_Y$ at positive gate voltages observed experimentally in Figure 2e.

## Normalised Berry curvature integral and comparison with experiment

Combining Equations S4, S7, and S10, the gate-dependent normalised Berry curvature integral is:

$$\Omega_{norm}(V_G) = \frac{\Delta k / k_F^2(V_G)}{\Delta k / (k_F^0)^2} = \frac{(k_F^0)^2}{k_F^2(V_G)} \qquad (S14)$$

This serves as a direct, electrically tunable measure of the quantum geometry sampled by the photoexcited carriers at the gate-controlled Fermi energy. The experimental proxy for $\Omega_{norm}(V_G)$ is the normalised peak $E_Y$ amplitude, since $E_Y \propto J_Y \propto \Omega_{int}$ (Equation S2). Note that $\Delta k$ cancels in the normalised ratio, so the result is independent of the absolute Weyl node separation. The experimental proxy for $\Omega_{norm}(V_G)$ is the normalised peak $E_Y$ amplitude, since $E_Y \propto J_Y \propto \Omega_{int}$ (Equation S2). Figure 1(d) shows the theoretical curve $\Omega_{norm}(V_G)$ alongside the normalised experimental $E_Y$ values across the full $\pm 90$ V gate voltage range. The quantitative agreement confirms that electrostatic Fermi pocket reshaping directly and continuously controls the Berry curvature integral sampled by the photoexcited carriers.